# Phase-change silicon as an ultrafast active photonic platform

Letian Wang*[1], Matthew Eliceiri[1], Yang Deng[2], Yoonsoo Rho[1], Wan Shou[3], Heng Pan[3], Jie Yao[2], Costas P. Grigoropoulos*[1]

[1] Laser Thermal Laboratory, Department of Mechanical Engineering, University of California, Berkeley, California, 94720, United States.
[2] Department of Material Science and Engineering, University of California, Berkeley, California, 94720, United States.
[3] Department of Mechanical and Aerospace Engineering, Missouri University of Science and Technology, Rolla, Missouri 65401, United States.

*Correspondence to: letianwang@berkeley.edu; cgrigoro@berkeley.edu

**Phase change material (PCM) features distinct optical or electronic properties between amorphous and crystalline states. Recently, it starts to play a key role in the emerging photonic applications like optoelectronic display[1], dynamic wavefront control[2], on-chip photonic memory[3] and computation[4,5]. However, current PCMs do not refract effectively at visible wavelengths[6] and suffer from deformation[7] and decomposition[7,8], limiting the repeatability and vast visible wavelength applications. Silicon as the fundamental material for electronics and photonics, has never been considered as phase change material, due to its ultrafast crystallization kinetics[9]. Here we show the striking fact that nanoscale silicon domains can be reversibly crystallized and amorphized under nanosecond laser pulses. For a typical disk resonator, it also provides a 25% non-volatile modulation at nanosecond time scale. We further show proof-of-concept experiments that such attributes could enable ultra-high resolution dielectric color display and dynamic visible wavefront control.**

**Main Text:**

Active photonics is the key to active wavefront control[2,10], structural color[1,11,12], optical memory[3] and computing[4,5,13]. The core of these applications is to develop fast active modulation of the sub-wavelength photonic element. Methods using phase change materials[5,14,15], electrical gating[16,17], liquid crystal[12,18] and thermal-optic effect[13,19], etc. have been studied to realize this goal. Among various methods, Phase Change Material (PCM) demonstrate a pivotal role[6] as it offers pixel-level modulation[2] and potentially ultrafast response[20]. Ternary chalcogenide glasses, such as GeSbTe (GST) constitute a prominent family of PCMs. They originate from optical storage and are now applied in active wavefront control[2,10], on-chip memory[3], and computation[4,5]. However, such glasses are not fully complementary metal-oxide-semiconductor (CMOS) compatible and require maintaining stoichiometry for consistent properties[8]. Even though reversible optical writing on GST thin films[2] is demonstrated, the same attempt is not successful on nanoscale elements [7], mainly due to the geometry deformation and chemical decomposition upon melting. Most importantly, they present a low refractive index and high loss in the visible wavelength, especially for the crystalline phase[3]. Consequently, active nanoresonators working effectively in the visible wavelengths are still not achieved by any PCMs.

In contrast, silicon as the fundamental material for CMOS technology features high refractive index and relatively low loss in the visible wavelength range. Furthermore, Si nanostructures present strong electrical and magnetic multipolar resonance[21] and various non-linear properties[22,23]. Hence, both crystalline and amorphous resonators have been used as meta-atoms for optical metasurfaces[21,24–26]. Nonetheless, silicon has never been considered as a phase change material as its fast nucleation and crystallization kinetics only permit amorphization at ultrafast quench rate and at the skin layer of a silicon wafer [27]. The photonic and electronic properties associated with crystallinity variation [28] may provide completely new opportunities to silicon-based nanophotonic devices[21].

Here, we present the concept of laser-induced reversible phase transformation of Si nanostructures in Fig. 1a. Arrays of amorphous Si (a-Si) nanodisks and nanoribbons were patterned from a 30nm amorphous thin film on fused silica, whose characteristic lengths range from 200 to 1500nm. The fabrication processes can be found in Methods and Supplementary Information S1. The forward phase transformation stands for the crystallization, which can be induced by either CW[29] or nanosecond laser[30]. Though not previously observed in nanostructures, the nanosecond laser-induced super-lateral growth[30] could provide fast crystallization speed, which is experimentally probed in this work. On the other hand, the CW laser crystallization features a wider laser power window, performing solid phase or nucleation based crystallization[29]. The CW laser crystallized nanodisks shown in Fig 1b, region A, are in the poly-crystalline state (c-Si) and present a green resonating color. Via backward phase transformation, c-Si nanodisks are amorphized by a 13ns laser pulse, leaving a red mark (region B, Fig.1b) on the green canvas. Those nanodisks share a "low density amorphous"(LDA) broad Raman peak at 450 cm$^{-1}$(Fig.1d). In Fig. 1c, additional CW laser crystallization resets the crystallinity and photonic resonance (region C, Fig.1c), indicating the initiation of next phase change cycle.

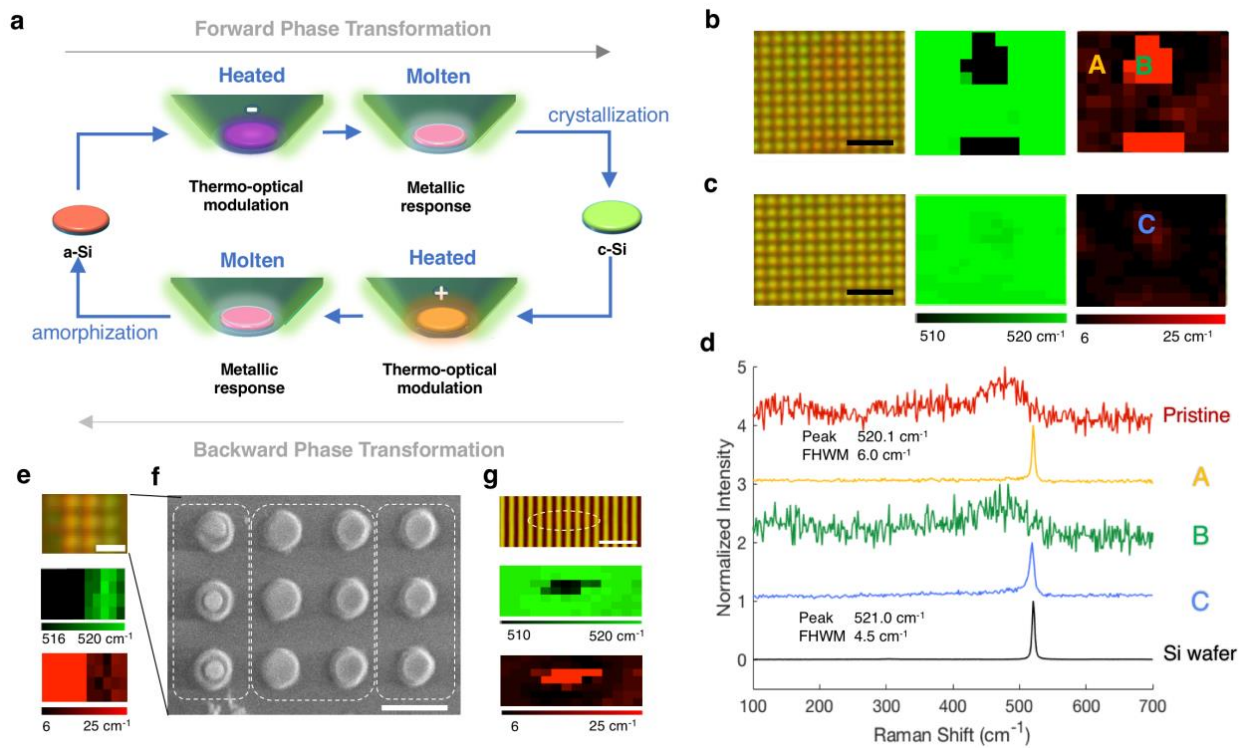

**Figure 1 Reversible phase transformations of silicon nanostructures.** (**a**) The schematics of the transient optical states during phase transformations. The "-" and "+" signs are the indicators of the opposite thermo-optical effects based on the crystallinity. (**b**) 210nm in-diameter polycrystalline silicon nanodisks amorphized by a nanosecond laser pulse, characterized with bright field optical image, mapping of Raman peak position and peak width(from left to right). (**c**) Amorphized nanodisks in (b) are "erased" with CW laser-induced crystallization. (**d**) Raman spectra of the labeled regions in b-c as well as pristine amorphous nanodisk and single-crystal silicon wafer. (**e**) Geometry pinning on 220nm nanodisk array characterized with the optical image and Raman mapping of peak position and width(from top to bottom). Bright field image contains dewetted disks (deep green), amorphized nanodisk (orange) and crystalline nanodisk (light green). (**f**) SEM image for e, with white dashed lines labeling aforementioned three regions. (**g**) Location-selective amorphization of 500nm wide nanoribbons. The laser-irradiated area is labeled with dashed lines. The scale bars for b-c are 2 μm, for e-f are 500nm and for g is 4 μm.

Amorphization can be obtained on a wide range of arbitrarily shaped nanostructures without deformation. As we can see in Fig. 1f, in contrast to the deformed dewetted nanodots, the morphology and shape of the amorphized nanodisks remain almost identical to crystalline nanodisks despite a slight edge rounding observed in nanodisks proximal to the dewetted dots. The transmission spectra of a-Si, c-Si and quenched amorphous Si (q-Si) confirm the optical resonance is also retained after the complete reversible phase change cycle (Supplementary Information S2). Silicon's unique high work of adhesion to oxides (3 times of gold[31]) thermodynamically prevents dewetting (Supplementary Information S3). Such geometry pinning effect eliminated the need for capping layers and therefore preserved the maximum optical contrast between the resonator and air. The amorphization and the geometry pinning applies to asymmetrical structures and large lateral sizes (Supplementary Information S2). Our reports show the maximum thickness for the amorphization to happen can be at least 100nm[32]. Furthermore, location-selective amorphization within one nanoribbon is achieved(Fig. 1g). The above results support the argument that the viable geometry for amorphization is governed by a critical local cooling rate[32]. The universal amorphization hence leads to wide applicability of reversible phase transformation on functional photonic structures.

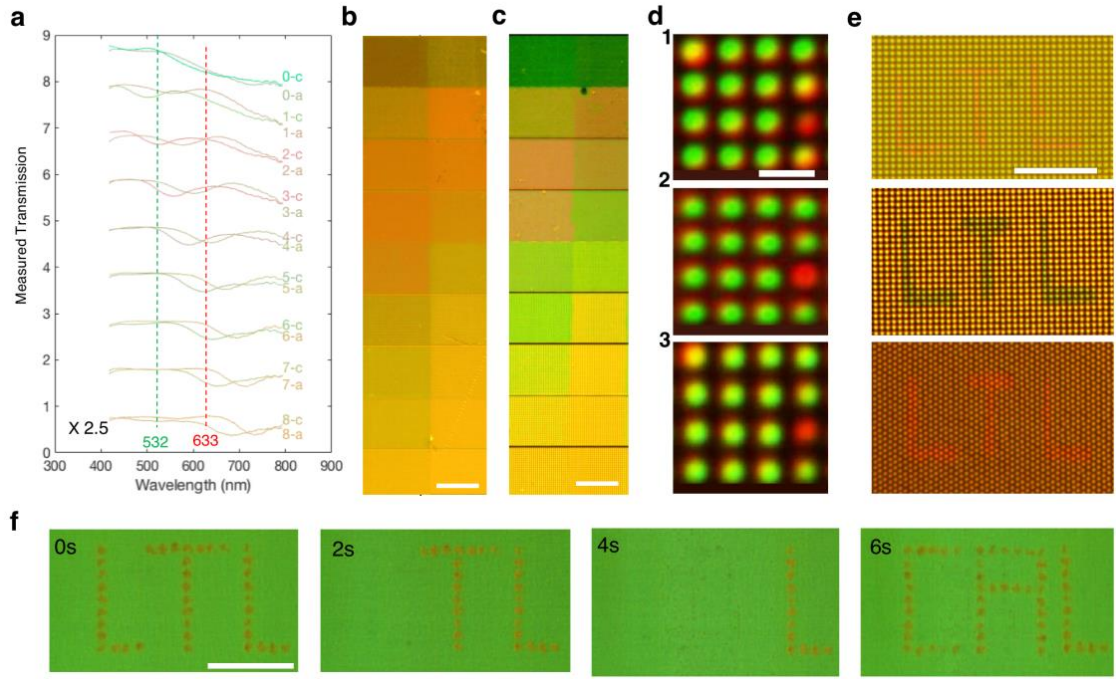

**Figure 2 Phase-dependent resonance shift, pixel-addressable active modulation and a dielectric display.** (**a**) The optical transmission spectra of the nanodisk arrays with different sizes and crystallinity. The diameters for number "0" to"8" nanodisks are 190nm, 215nm, 260nm, 285nm, 310nm, 335nm, 370nm, 395nm, and 420nm respectively. The crystalline and amorphous phases are labeled as (-c) and (-a). The signal is multiplied by 2.5 and stacked for better visualization. (**b-c**) The optical (b)bright field and (c)dark field images of the crystalline (left) and amorphous (right) nanodisk arrays, with diameter ranged from number "0" to"8" from top to bottom. (**d**) 2D diffraction-limited pixel-addressable encoding. The 4-by-4 crystalline nanodisk array is originally resonating green color, and red dot (3,4) is a dewetted disk for reference. First, amorphization turns disk (1,1), (2,4) and (4,4) into yellow color. Then the crystallization erases dot (1,1), (2,4) and (4,4) into green color. Lastly, re-amorphization addressed the same disk (1,1) again. (**e**) Diffraction limited laser printing of "LTL" letters on different amorphous disk arrays, whose diameter/pitch are 310nm/700nm, 335nm/750nm, 285nm/650 nm (hcp) array, from top to bottom. (**f**) A dynamic color display showing the characters' transition from "LTL" to "CAL." The scale bars for b-f are 20 μm, 20 μm, 1 μm, 8 μm, and 20 μm.

Based on the reversible resonance shift, we demonstrate pixel level active modulation on visible photonic arrays. The optical resonances of silicon nanostructures span broad visible spectra (Fig. 2a-c), leading to wide color palettes[33]. Upon amorphization, the refractive index in the visible range is reduced and induces redshifts of multipolar resonances (Fig.2a and Supplementary Information S4). Enabled by high N.A. objective lens, we show reversible encoding of digital information at selected pixels on 700nm pitched nanoresonator arrays (Fig. 2d), which is the first demonstration of near-diffraction-limit pixel level active modulation of a photonic array. With high-resolution patterning and resonating printing [34], the minimum optical addressable resolution is further reduced to 400nm per pixel (Supplementary Information S5), paving the way to pixel-rewritable metasurfaces working at wavelengths below 800nm. Corresponding to the concept of plasmonic displays[11], we show high-resolution dielectric display of "LTL" letters (Fig. 2e), which offers a display resolution up to 63,500 PPI, two orders of magnitude higher than today's mainstream electronic displays. As a proof-of-concept experiment, we further demonstrate a dynamic dielectric display showing text "LTL" transitioning to "CAL" (Fig. 2f). Despite that only 0.5 Hz refresh rate is obtained on a motion stage, a high refresh rate display can be achieved through laser scanning digital light projection (DLP).

Optical reflection probing characterized the time scale of nanosecond laser-induced amorphization and crystallization. The 633nm probing CW laser and the 532nm pump nanosecond laser are aligned tightly to a group of nanoresonators(Fig. 3a). From Fig. 3c-d, a distinctive high reflective state ($\beta$) signifies the High-Density Liquid phase[35] during the laser-induced phase transformations. The drop of high reflection state leads to a different level of reflection ($\gamma$) compared to the original level($\alpha$), which corresponds to phase-dependent Mie resonance. FDTD simulations on the silicon disk's $\alpha$, $\beta$ and $\gamma$ are provided in Supplementary Information S4. Indicated by the prolonged high reflection level(Fig. 3d), we concluded the amorphization requires full-melting. Together with Supplementary Information S6, we know the crystallization is under near-complete melting fluence. Given the short melt duration of crystallization(Fig.3c), all the evidence leads to a super-lateral growth based crystallization mechanism[30], where non-melting residual crystals serve as the seeds for crystal growth. Both forward and backward phase transformations conclude within 100ns, including the heating, melting, solidification and cooling processes. Consequently, the reversible switching frequency reaches 5MHz, which is comparable to industry start-of-the-art[36]. Up to 25% non-volatile modulation can be provided in the probed nanoresonators (Fig.3c-d and f, $\gamma/\alpha$). The contrast is comparable to that of GST based on-chip photonic memories[3]. With the proper design of nanoresonator geometries, the optical contrast can be further maximized.

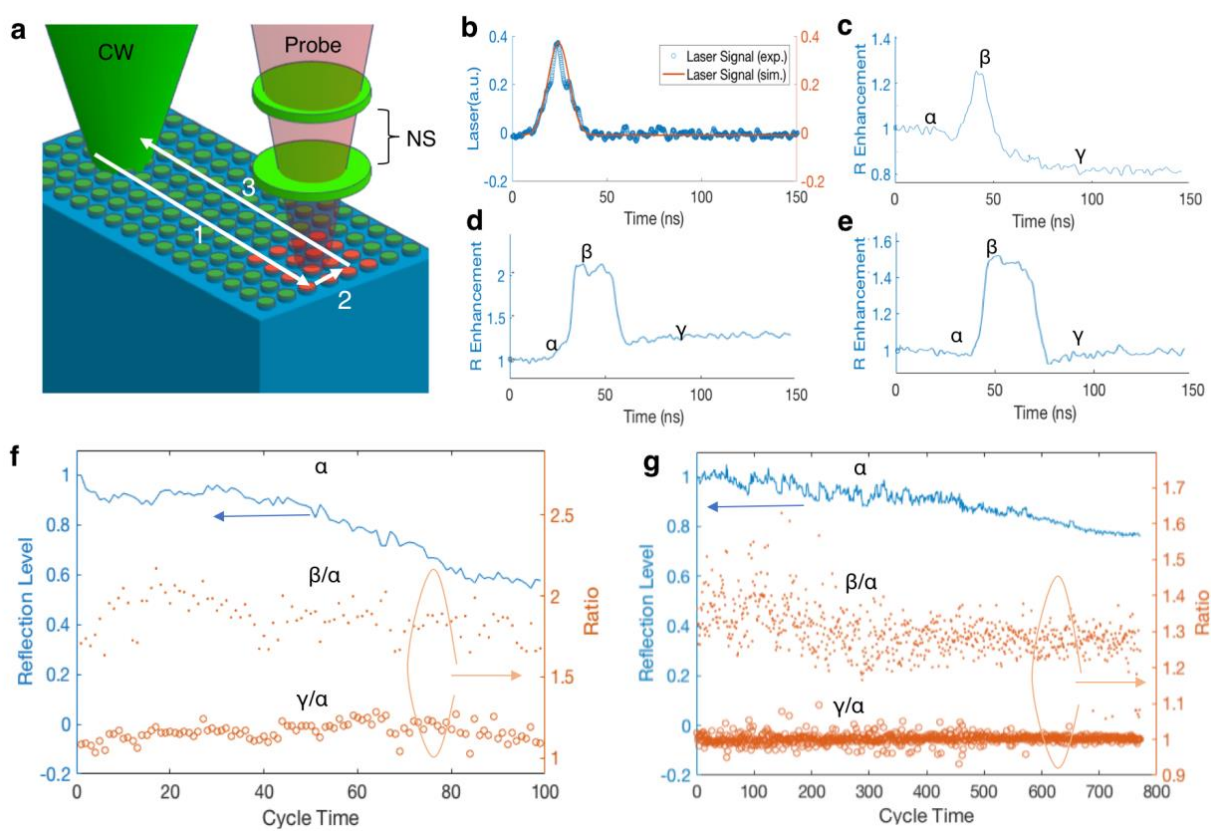

**Figure 3 Characterization of ultrafast non-volatile reversible phase change and repeatable cycle lifetime**. (**a**) Schematics of the optical reflection probing and repeated cycle lifetime. 1->2->3 stands for the path of additional CW laser scanning for repeatable lifetime measurement. (**b**) Probed nanosecond pulsed laser signals and gaussian fitted laser pulse. (**c-e**) Probed transient reflection of forward (a-Si to c-Si, **c**), backward(c-Si to a-Si, **d**) and neutral (a-Si to a-Si, **e**) phase transformations. α, β, γ stand for the levels of reflection signals before-transformation, fully molten and after-transformed states. All signals are normalized with α. (**f-g**) Repeatable cycle lifetime of 210 nm diameter bare resonators with (j) forward and (k) neutral phase transformations.

The cap-less reversible cycle lifetime is characterized through recording the α, β, γ parameters (see the caption of Fig.3d-e) during consecutive phase transformations. The absolute reflectance (α) is selected as the performance indicator for evaluating the lifetime of the resonator. For complete reversible phase transformations, CW laser scanning is applied to ensure more uniform quality of crystallization than nanosecond laser pulses. The nanodisks yield 50 cycles with only 10% degradation in α (0.5dB) but fully dewet after 80 cycles (Fig. 3f). Besides the reversible phase changes, we probed the lifetime of repeated neutral (a-Si to a-Si) transformations (Fig. 3e). Figure. 3g shows that the disks withstand ~400 such cycles with only 10% degradation in α (0.5dB) but fully dewet after 700 cycles. The decoupling of crystallization points out significant degradation effect are from non-optimized crystallization power, and promise the further improvement of the cycle lifetime. In Fig. 3e-f, β/α equals to 2 (or 1.3 depending on the initial crystallinity) ensures the resonators passed melting threshold. The ratio γ/α describes the direction of phase transformation, i.e. backward (>1) or neutral (=1) phase transformation. Details about characterization can be found in Supplementary Information S7. Note that if high optical contrast is not intended, degradation and deformation can be fully suppressed with additional capping layers.

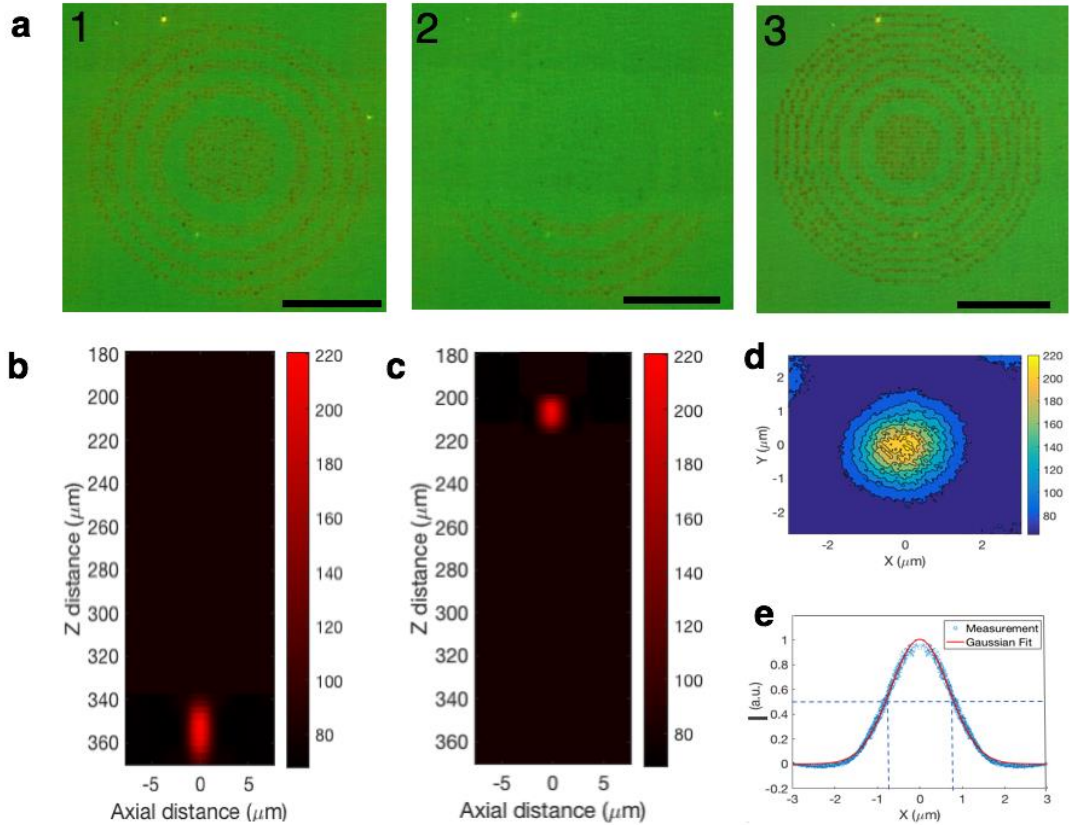

**Figure 4 Active visible Fresnel zone plates with the on-demand tuning of focal lengths.** (**a**) Dark field images of (1) writing f=400 μm, (2) erasing and (3) rewriting f=240 μm FZPs on the same canvas and location. The scale bar is 30μm. (**b-c**) Measured spatial intensity distribution near focal planes of printed FZPs from A. The actual focal length is measured to be 353 μm and 207 μm respectively. The higher orders are not shown in the plot. (**d**) Focal spot of 1$^{st}$ FZP characterized with CCD intensity image. (**e**)The line plot of radially accumulated intensity and Gaussian curve fitting.

Lastly, we show a proof-of-concept experiment of an active optical element in the visible wavelength. On the exact same place, amplitude-based Fresnel Zone Plates (FZPs) are actively printed with different focal lengths (Fig. 4a-c). The measured cross-section intensity distribution clearly showed the FZP's focal spot has shifted from 200 μm to 350 μm with a slight deviation from designed values (240 μm and 400 μm). The deviation is likely due to the printing pixel limitation and imperfect fabrication[37]. The laser printed FZPs achieved a diffraction-limited ($\lambda$/2NA, $\lambda$=633nm, NA=100/2f=0.21), full-width at the half-maximum focal spot of about 1.5 μm (Fig. 4d-e). Setups and characterization of the focal plane are described in Supplementary Information S8. This demonstration differs from the previous reports[2] in that it is based on individual resonators, works at visible wavelengths, and can be easily cycled multiple times. If the light field is delivered through spatial light modulation (SLM) , such components can be arbitrarily written [37] and printed at nanosecond time scale. Besides the optical wavefront control, rewritable metasurfaces can also be readily applied to holography and encryption[38].

The phase change silicon as an ultrafast active photonic platform further opens considerable opportunities. Given reported femtosecond(fs) laser-induced crystallization[39] and amorphization[27] effects on silicon, we anticipate the introduction of fs laser on silicon nanoresonators can further improve the switching frequency up to 100MHz (Supplementary Information S9). With the controlled number of fs laser pulses or cooling time, a multilevel response can be obtained with a combination of crystalline and amorphous phases[39,40]. As the key amorphization requires only nanosecond modulation, it is possible to use electrical pulses to phase change an ultra-small volume (0.01 μm$^3$). Our analysis shows the energy consumption is at the same order of applying GST materials (Supplementary Information S10). Beyond the resonance shift, silicon has crystallinity dependent thermo-optical and non-linear properties[40], which are intriguing for implementing active tuning functionalities. Ultimately, Ge, GaP and GaAs share similar potential as phase change materials[41] and may generate novel functional active photonics[42].

## Methods

**Nanofabrication.** 30 nm amorphous Si (a-Si) is first deposited on 500 μm fused silica (HOYA) substrate via Low-Pressure Chemical Vapor Deposition (LPCVD) at 500°C. The deposited film is then patterned with standard photolithography (ASML 5500/300) and reactive ion etching (RIE, $Cl_2$, HBr and $O_2$). Over-etch during RIE process leaves silicon sitting on a 50nm thick oxide mesa. The schematics of the process flow can be found in Supplementary Information. S1. Through the transfer-based fabrication method[26], we obtained 40nm single crystal silicon membrane on fused silica. Then asymmetric geometries and the 400nm period nanodisk arrays are patterned through Electron Beam lithography (Crestec) and same RIE etching.

**Laser Processing and Transient Reflection Probing.** A high power Nd:YAG pulsed laser (Polaris III-10, New Wave Laser, 532nm, 13 ns measured pulse duration) is employed to amorphize the nanodisks. The RMS energy fluctuation was measured to be about 4%. A CW laser (Lighthouse Sprout-C 4W, 532nm) is used to crystallize large areas of the a-Si nanostructures to c-Si. Mitutoyo APO 10X and 5X objective lenses are used for generating large area amorphization and crystallization. Additionally, Olympus MS Plan 100X (NA 0.95) is used for demonstrating pixel-addressable modulation and single disk transient probing. For transient probing, a HeNe Laser (JDSU, 632.8nm, 15mW) is employed to probe the reflection. The reflection signal is probed through a high-speed avalanche photodetector (Thorlabs APD, 0-400MHz, 1ns rise time). Details on the characterization of cycle lifetime are included in the Supplementary Information S10. The printing of FZP is achieved with Mitutoyo APO 50X.

**Characterization on Optical Response, Morphology and Raman Spectra.** An optical microscope (Olympus BX60) and CMOS camera (AmScope MU300) have been utilized to characterize the optical bright field and dark field images. Scanning Electron Microscopy (Quanta FEI) is utilized to characterize the morphology nanodisks and nanodots. No metal coating is applied. A Renishaw inVia Micro Raman System with a 532nm laser excitation source and a 100nm-resolution motion stage are used for Raman spectra mapping. The mapped spectra are analyzed with the built-in WiRE 3.3 software. Spectra curves are fitted with combined Gaussian and Lorentzian profiles with the peak centered between 480-522 cm$^{-1}$. The static transmission spectra are measured through a 100X objective lens (Olympus MS Plan, NA0.95) with a stabilized light source (Thorlabs SLS201), terminating at a spectrometer (Ocean Optics USB4000). The FZP focal spot intensity distribution is probed with a home-built transmission setup with Mitutoyo 50X objective lens, a CCD camera (Amscope MD1000) and a high-precision translation stage (Thorlabs Nanomax 302). The setup details can be found in Supplementary Information S11.

**Optical Simulation.** Commercial simulation software (Lumerical FDTD[43]) based on the finite-difference-time-domain method is applied to simulate the optical reflection for different states of nanodisks. A 30nm thick nanodisk with 200nm diameter is placed on top of a 50nm oxide mesa and then on 5μm quartz with top-down plane-wave irradiation. Refractive indices of crystalline Si (c-Si) and amorphous Si (a-Si) properties are selected from built-in models and literature[44] respectively. The temperature-dependent refractive index and molten silicon properties are taken from literature[45].

**Thermal and Crystallization Simulation**. As no dewetting and deformation processes are observed for the nanodisks, a fixed geometry finite difference simulation is carried out using the same simulator on the previous reports[32,46,47]. A single Si nanodisk is placed on the oxide mesa and then on an oxide substrate. The nanodisk and mesa are modeled as cylinders with the same dimensions as optical simulations(200nm in-diameter, 30nm thick) with 5nm grid mesh. The x- and y-direction boundaries are set to be adiabatic, which is valid given periodic arrangement of the nanodisks. The thermal conduction to the air and radiation to the ambient are neglected. The simulated silica substrate thickness is set to be 2 times the vertical heat diffusion length. The simulator is based on the quasi-steady state classical nucleation theories.

**References:**

1. Hosseini, P., Wright, C. D. & Bhaskaran, H. An optoelectronic framework enabled by low-dimensional phase-change films. *Nature* **511**, 206–211 (2014).
2. Wang, Q. *et al.* Optically reconfigurable metasurfaces and photonic devices based on phase change materials. *Nat. Photonics* **10**, 60–65 (2016).
3. Rios, C. *et al.* Integrated all-photonic non-volatile multi-level memory. *Nat. Photonics* **9**, 725–732 (2015).
4. Cheng, Z., Ríos, C., Pernice, W. H. P., Wright, C. D. & Bhaskaran, H. On-chip photonic synapse. *Sci. Adv.* **3**, e1700160 (2017).
5. Cheng, Z. *et al.* Device-Level Photonic Memories and Logic Applications Using Phase-Change Materials. *Adv. Mater.* **30**, (2018).
6. Wuttig, M., Bhaskaran, H. & Taubner, T. Phase-change materials for non-volatile photonic applications. *Nat. Photonics* **11**, 465–476 (2017).
7. Karvounis, A., Gholipour, B., MacDonald, K. F. & Zheludev, N. I. All-dielectric phase-change reconfigurable metasurface. *Appl. Phys. Lett.* **109**, (2016).
8. Salinga, M. *et al.* Monatomic phase change memory. *Nat. Mater.* **17**, 681–685 (2018).
9. Bhat, M. H. *et al.* Vitrification of a monatomic metallic liquid. *Nature* **448**, 787–790 (2007).
10. Yin, X. *et al.* Beam switching and bifocal zoom lensing using active plasmonic metasurfaces. *Light Sci. Appl.* **6**, e17016 (2017).
11. Duan, X., Kamin, S. & Liu, N. Dynamic plasmonic colour display. *Nat. Commun.* **8**, 14606 (2017).
12. Franklin, D., Frank, R., Wu, S. T. & Chanda, D. Actively addressed single pixel full-colour plasmonic display. *Nat. Commun.* **8**, 1–10 (2017).
13. Shen, Y. *et al.* Deep learning with coherent nanophotonic circuits. *Nat. Photonics* **11**, 441–446 (2017).
14. Oh, Y. *et al.* Plasmonic Periodic Nanodot Arrays via Laser Interference Lithography for Organic Photovoltaic Cells with >10% Efficiency. *ACS Nano* **10**, 10143–10151 (2016).
15. Dong, K. *et al.* A Lithography-Free and Field-Programmable Photonic Metacanvas. *Adv. Mater.* **30**, 1703878 (2018).
16. Huang, Y. W. *et al.* Gate-Tunable Conducting Oxide Metasurfaces. *Nano Lett.* **16**, 5319–5325 (2016).
17. Kim, S. J. & Brongersma, M. L. Active flat optics using a guided mode resonance. *Opt. Lett.* **42**, 5 (2017).
18. Sautter, J. *et al.* Active tuning of all-dielectric metasurfaces. *ACS Nano* **9**, 4308–4315 (2015).
19. Horie, Y., Arbabi, A., Arbabi, E., Kamali, S. M. & Faraon, A. High-Speed, Phase-Dominant Spatial Light Modulation with Silicon-Based Active Resonant Antennas. *ACS Photonics* **5**, 1711–1717 (2018).
20. Michel, A. K. U. *et al.* Reversible Optical Switching of Infrared Antenna Resonances with Ultrathin Phase-Change Layers Using Femtosecond Laser Pulses. *ACS Photonics* **1**, 833–839 (2014).
21. Staude, I. & Schilling, J. Metamaterial-inspired silicon nanophotonics. *Nat. Photonics* **11**, 274–284 (2017).
22. Shcherbakov, M. R. *et al.* Enhanced third-harmonic generation in silicon nanoparticles

23. Liu, S. *et al.* Resonantly Enhanced Second-Harmonic Generation Using III-V Semiconductor All-Dielectric Metasurfaces. *Nano Lett.* **16**, 5426–5432 (2016).
24. Lin, D., Fan, P., Hasman, E. & Brongersma, M. L. Dielectric gradient metasurface optical elements. *Science* **345**, 298–302 (2014).
25. Arbabi, A., Horie, Y., Bagheri, M. & Faraon, A. Dielectric metasurfaces for complete control of phase and polarization with subwavelength spatial resolution and high transmission. *Nat. Nanotechnol.* **10**, 937–943 (2015).
26. Deng, Y. *et al.* All-Silicon Broadband Ultraviolet Metasurfaces. *Adv. Mater.* **0**, 1802632 (2018).
27. Izawa, Y. *et al.* Ultrathin amorphization of single-crystal silicon by ultraviolet femtosecond laser pulse irradiation. *J. Appl. Phys.* **105**, (2009).
28. Han, D. *et al.* Optical and electronic properties of microcrystalline silicon as a function of microcrystallinity. *J. Appl. Phys.* **87**, 1882–1888 (2000).
29. Jin, S. *et al.* Lateral grain growth of amorphous silicon films with wide thickness range by blue laser annealing and application to high performance poly-Si TFTs. *IEEE Electron Device Lett.* **37**, 291–294 (2016).
30. Im, J. S., Kim, H. J. & Thompson, M. O. Phase transformation mechanisms involved in excimer laser crystallization of amorphous silicon films. *Appl. Phys. Lett.* **63**, 1969–1971 (1993).
31. Sangiorgi, R., Muolo, M. L., Chatain, D. & Eustathopoulos, N. Wettability and Work of Adhesion of Nonreactive Liquid Metals on Silica. *J. Am. Ceram. Soc.* **71**, 742–748 (1988).
32. Wang, L. *et al.* Programming Nanoparticles in Multiscale: Optically Modulated Assembly and Phase Switching of Silicon Nanoparticle Array. *ACS Nano* **12**, 2231–2241 (2018).
33. Flauraud, V., Reyes, M., Paniagua-Dominguez, R., Kuznetsov, A. I. & Brugger, J. Silicon nanostructures for bright field full color prints. *Acs Photonics* **4**, 1913–1919 (2017).
34. Zhu, X., Yan, W., Levy, U., Mortensen, N. A. & Kristensen, A. Resonant laser printing of structural colors on high-index dielectric metasurfaces. *Sci. Adv.* **3**, (2017).
35. Beye, M., Sorgenfrei, F., Schlotter, W. F., Wurth, W. & Fohlisch, A. The liquid-liquid phase transition in silicon revealed by snapshots of valence electrons. *Proc. Natl. Acad. Sci.* **107**, 16772–16776 (2010).
36. AGILTRON. High Speed Optical Switches - NanoSpeed. Available at: https://agiltron.com/category/fiber-optic-switches/nanospeed-fiber-optical-switches/.
37. Carstensen, M. S. *et al.* Holographic Resonant Laser Printing of Metasurfaces Using Plasmonic Template. *ACS Photonics* **5**, 1665–1670 (2018).
38. Li, J. *et al.* Addressable metasurfaces for dynamic holography and optical information encryption. *Sci. Adv.* **4**, eaar6768-eaar6768 (2018).
39. Zywietz, U., Evlyukhin, A. B., Reinhardt, C. & Chichkov, B. N. Laser printing of silicon nanoparticles with resonant optical electric and magnetic responses. *Nat. Commun.* **5**, 3402 (2014).
40. Makarov, S., Kolotova, L., Starikov, S., Zywietz, U. & Chichkov, B. Resonant silicon nanoparticles with controllable crystalline states and nonlinear optical responses. *Nanoscale* **10**, 11403–11409 (2018).
41. Cullis, A. G., Webber, H. C. & Chew, N. G. Amorphization of germanium, gallium phosphide, and gallium arsenide by laser quenching from the melt. *Appl. Phys. Lett.* **42**, 875–877 (1983).

42. Ha, S. T. *et al.* Directional lasing in resonant semiconductor nanoantenna arrays. *Nat. Nanotechnol.* **13**, 1042–1047 (2018).
43. Lumerical Solutions, Inc.
44. Demichelis, F. *et al.* Optical properties of hydrogenated amorphous silicon. *J. Appl. Phys.* **59**, 611–618 (1986).
45. Yavas, O. *et al.* Temperature dependence of optical properties for amorphous silicon at wavelengths of 632.8 and 752 nm. *Opt. Lett.* **18**, 540–542 (1993).
46. Xiang, B. *et al.* In situ TEM near-field optical probing of nanoscale silicon crystallization. *Nano Lett.* **12**, 2524–9 (2012).
47. In, J. Bin *et al.* Generation of single-crystalline domain in nano-scale silicon pillars by near-field short pulsed laser. *Appl. Phys. A* **114**, 277–285 (2014).

**Acknowledgments**

The work is supported by the collaborative National Science Foundation Grant CMMI- 1363392 to C.G. and H.P., Y.D and J.Y. acknowledge the support by the National Science Foundation DMR-1555336. The authors would like to thank Prof. Paul McMillan at University College London for helpful discussion. The nanofabrication and SEM were carried out at the Marvell Nanofabrication Laboratory and the California Institute of Quantitative Bioscience (QB3) in UC Berkeley. L.W. would like to thank Chris Zhao and Zeying Ren's input on the nanofabrication. Lastly, the authors appreciate Dr. Jung Bin In's contribution to developing the simulator

**Author Contributions**

L.W. conceived the idea, designed and carried out most experiments, and all the numerical analysis. M.E. contributed to the laser processing and characterization of reversible experiments. Y.R. contributed to the optical setup for in-situ probing. Y.D. and J.Y contributed to the electron beam lithography fabrication and optical analysis. W.S. and H.P. offered formal analysis on phase transformation. C.G. supervised and supported the work;

**Author Information**

The authors declare competing financial interest. Correspondence and requests for materials should be addressed to L.W.(letianwang@berkeley.edu) and C.G. (cgrigoro@berkeley.edu).

**Additional Information**
**Supplementary information** is available for this paper at https://doi.org/XX.
**Reprints and permissions information** is available at XXXX.